\documentclass{icrc2009}

\usepackage{graphicx}   
\usepackage[caption=false]{caption}    
\usepackage[font=footnotesize]{subfig} 
\usepackage{fixltx2e}
\usepackage{url}

\newcommand{\shorttitle}[1]%
{\markboth{Proceedings of the 31\MakeLowercase{$^{st}$} ICRC, {\L}\'{o}d\'{z} 2009}{#1} }

\newcommand{\pubjournal}[6] {#1, #2 {\bf #3}, #4 (#5).}

\begin{document}
\title{A Galactic Cosmic-Ray Database}

\author{\IEEEauthorblockN{A.W. Strong \IEEEauthorrefmark{1}, 
			  I.V. Moskalenko \IEEEauthorrefmark{2}    
   }
                            \\
\IEEEauthorblockA{\IEEEauthorrefmark{1}Max-Planck-Institut f\"ur extraterrestrische Physik, Garching, Germany}
\IEEEauthorblockA{\IEEEauthorrefmark{2}Hansen Experimental Physics Laboratory and Kavli Institute for Particle  Astrophysics and Cosmology,\\ Stanford University, Stanford, CA 94304, USA}}

\shorttitle{Strong and Moskalenko: Cosmic-Ray Database}

\maketitle

\begin{abstract}
Despite a century  of cosmic-ray measurements there seems to have been no
attempt  to collect these data systematically for use by the
CR community.
The result is that everyone makes their own collection as required, a large
duplication of effort.
Therefore we have started a project to place published Galactic CR measurements in
a database available online.
It currently addresses energies up to 100 TeV, including elemental
spectra, secondary/primary nuclei ratios, antiprotons, electrons and  positrons.
It is updated regularly as data appears in the literature.  It is supported by access software.
The community is encouraged to participate by providing data, pointing out
errors or omissions, and suggestions for improvements.
\end{abstract}


\section{Introduction}
While great effort is expended on the experiments on which cosmic-ray science is based,
not much attention has been given  (as far as we know)  to the collection of such data in a convenient format.
The result is that everyone makes their own compilation, which seems to us a waste of effort,
also leading to possibly incomplete and/or error-prone values.
Therefore we decided to initiate a simple database project for GCR measurements, with the idea that it would grow to become of real use to the  community.
The project actually began while writing  the review \cite{AnnRev57}, which included a representative set of plots from the literature, but
not in the uniform style which is now possible with this database.

We restrict our data to Galactic CR at energies below about $10^{14}$ eV, excluding air-shower data (which is beyond our scope but could be considered if supported by the relevant experts). It includes nuclei - elemental and isotopic, electrons, positrons and antiprotons. Both flux spectra and ratios like B/C are supported.

The database is developed in association with the  GALPROP code\footnote{http://galprop.stanford.edu} for which
more details can be found in \cite{SM98,SMR00,Moskalenko2003,SMR04,SMRRD04,Strong2009}.

\section{Data collection}
In many cases data are provided in published papers in the form of tables. In this case only a conversion of units to a standard form is required.
In other cases only plots are presented, in which case the authors are contacted for a tabular version, which has  always been provided so far.
The reference is included in the database, including a note on the form in which the data were obtained.
The experiments currently represented with data include:
ACE-CRIS,  AMS01, ATIC1+2,  BESS,   BETS, CAPRICE, CRN,  Fermi-LAT,  HEAO3, HEAT, HESS,   IMAX, IMP, JACEE,  MUBEE, PAMELA,  PPB-BETS,
RUNJOB,   SANRIKU, SOKOL, TIGER, Voyager.

\section{Format}
The format is simple ASCII (i.e. it is not a database  in the technical sense).
Results are grouped according to the reference used, identified by a character string, while the experiment is identified by another string.
Entries contain the kinetic energy (per nucleon for nuclei), units,  a flag to indicate a flux or a ratio, error bars (upper and lower if appropriate), Z and A of the species, and comments.
Elements are flagged as all A for a given Z.  Up to 3 nuclei can be combined by a list of Z and A.
Some flexibility in the units is allowed ( MeV or GeV, cm$^{-2}$ or m$^{-2}$ ) for convenience of checking against the original data used,
 but these are  converted into  user-required units by the software interface provided.
The volume of data is easily accomodated in a small file\footnote{Published GCR data total only about 200~kB !}.

 \section{Software support}
A C++ class is provided which allows spectra to be extracted for a given CR element or isotope, combination of species, or electrons, positrons or antiprotons.
 For ratios, the `primary' and `secondary' species must be specified (including combinations like the sub-Fe isotopes).
 A Python interface has  also been provided\footnote{L.~Baldini,\\ http://www-glast.stanford.edu/cgi-bin/viewcvs/users/lbaldini/GCRdb},
an example of a user's contribution to the project.




\section{Management and community use}
The data are maintained on the website  http://www.mpe.mpg.de/$\sim$aws/propagate.html and via the
GALPROP website.
Updates are made as  required, when  new data becomes available or improvements are made. Changes are documented at the end of the file.
 The  forum on the GALPROP website includes a section related to this database.
The reaction so far from those who have made use of these data  has been enthusiastic.
We intend to continue to add data, relying on users' inputs to find errors, make suggestions for improvements and especially for new data.

%
%
%




\section{Future developments}
While the ASCII format is simple, more flexibility would be provided by a FITS table, and this is foreseen as an alternative in future.
Extension to higher energies will be considered. Suggestions are welcome.

\section{Acknowledgments}

I. V. Moskalenko  acknowledges support from NASA
grant NNX09AC15G.

\end{document}